%% file: main.tex
\documentclass[sigconf,screen]{acmart}

\AtBeginDocument{%
  \providecommand\BibTeX{{%
    \normalfont B\kern-0.5em{\scshape i\kern-0.25em b}\kern-0.8em\TeX}}}

\setcopyright{none}

\acmConference[CHI '24]{CHI Conference on Human Factors in Computing Systems}{May 11--16, 2024}{Honolulu, Hawaii}

\input{00-0packages}

\begin{document}

\title[Examining the Effects of Witnessing Post-Removal Explanations]{Bystanders of Online Moderation:\\ Examining the Effects of Witnessing Post-Removal Explanations}


\author{Shagun Jhaver}
\orcid{0000-0002-6728-7101}
\affiliation{%
  \institution{Rutgers University}
  \city{New Brunswick}
  \state{NJ}
  \country{USA}}
\email{shagun.jhaver@rutgers.edu}

\author{Himanshu Rathi}
\affiliation{%
  \institution{Rutgers University}
  \city{New Brunswick}
  \state{NJ}
  \country{USA}}
\email{hr393@scarletmail.rutgers.edu}

\author{Koustuv Saha}
\orcid{0000-0002-8872-2934}
\affiliation{%
 \institution{University of Illinois}
 \city{Urbana-Champaign}
 \state{IL}
 \country{USA}}
 \email{ksaha2@illinois.edu}

\renewcommand{\shortauthors}{Jhaver et al.}

\begin{abstract}
Prior research on transparency in content moderation has demonstrated the benefits of offering post-removal explanations to sanctioned users. In this paper, we examine whether the influence of such explanations transcends those who are moderated to the bystanders who witness such explanations. We conduct a quasi-experimental study on two popular Reddit communities (r/askreddit and r/science) by collecting their data spanning 13 months—a total of 85.5M posts made by 5.9M users. Our causal-inference analyses show that bystanders significantly increase their posting activity and interactivity levels as compared to their matched control set of users. 
In line with previous applications of Deterrence Theory on digital platforms, our findings highlight that understanding the rationales behind sanctions on another user significantly shapes observers’ behaviors.
We discuss the theoretical implications and design recommendations of this research, focusing on how investing more efforts in post-removal explanations can help build thriving online communities.
\end{abstract}

\ccsdesc[300]{Human-centered computing~Empirical studies in collaborative and social computing}
\ccsdesc[300]{Human-centered computing~Social media}

\keywords{content moderation, social media, transparency, causal-inference}


\maketitle

\input{00-body}

\begin{acks}
We thank Anish Gupta, Gayathri Ravipati, and Navreet Kaur for their work on this project. 

\end{acks}

\bibliographystyle{ACM-Reference-Format}
\bibliography{references_moderation,references_more,references_ks}

\end{document}
\endinput

%% file: 00-0packages.tex
\usepackage{capt-of}
\usepackage{balance,bm}
\usepackage{color, colortbl, xcolor}

\usepackage{url}
\usepackage{hyperref}
\hypersetup{
    colorlinks=true,
}

\usepackage{subcaption}
\usepackage{booktabs} 
\usepackage{graphicx}
\usepackage{textcomp}
\usepackage{color,soul}
\usepackage{bm}
\usepackage{multirow}
\usepackage{wrapfig}
\usepackage{balance}
\usepackage{graphicx}  

\usepackage{colortbl}
\usepackage{arydshln}
\usepackage [english]{babel}
\usepackage [autostyle, english = american]{csquotes}
\definecolor{linkColor}{RGB}{6,125,233}
\definecolor{green}{rgb}{0.0, 0.65, 0.31}
\definecolor{bleudefrance}{rgb}{0.19, 0.55, 0.91}
\definecolor{ceruleanblue}{rgb}{0.16, 0.32, 0.75}
\definecolor{grey}{HTML}{969696}
\definecolor{violet}{HTML}{756bb1}
\definecolor{dgrey}{HTML}{01665e}
\definecolor{lgrey}{HTML}{5ab4ac}
\definecolor{dgreen}{HTML}{005a32}
\definecolor{purple}{HTML}{ae017e}

\definecolor{editCol}{HTML}{000000}
\definecolor{maskCol}{HTML}{c51b7d}
\definecolor{lrColor}{HTML}{8856a7}
\definecolor{trColor}{HTML}{d01c8b}
\definecolor{ctColor}{HTML}{4dac26}
\definecolor{brickred}{HTML}{f03b20}
\definecolor{improveCol}{HTML}{253494}
\definecolor{worsenCol}{HTML}{d7191c}
\definecolor{DarkBlue}{HTML}{00008B}
\definecolor{mscolor}{HTML}{01665e}
\definecolor{nmscolor}{HTML}{bf812d}
\definecolor{lgreen}{HTML}{ccece6}
\definecolor{dolive}{HTML}{308014}

\usepackage{mathtools}

\usepackage{amsmath}
\usepackage{array}
\usepackage{xcolor}
\usepackage{arydshln}
\usepackage{siunitx}
\colorlet{tablerowcolor4}{gray!50} 

\newcommand*{\textlabel}[2]{%
  \edef\@currentlabel{#1}
  \phantomsection
  #1\label{#2}
}
\usepackage{tcolorbox}

\colorlet{tableheadcolor}{gray!25} 
\colorlet{tablerowcolor}{gray!15} 
\colorlet{tablerowcolor2}{gray!45} 
\colorlet{tablerowcolor3}{gray!25} 

\newcommand{\rowcollight}{\rowcolor{tablerowcolor3}} %

\newif{\ifhidecomments}
  \hidecommentsfalse 
\ifhidecomments
    \newcommand{\himanshu}[1]{}
    \newcommand{\koustuv}[1]{}
    \newcommand{\shagun}[1]{}
\else
    \newcommand{\himanshu}[1]{\textbf{\small\sffamily{\textcolor{orange}{[#1 -- Himanshu]}}}}
    \newcommand{\koustuv}[1]{\textbf{\small\sffamily{\textcolor{DarkBlue}{[#1 -- Koustuv]}}}}
    \newcommand{\shagun}[1]{\textbf{\small\sffamily{\textcolor{dgreen}{[#1 -- Shagun]}}}}
  \fi

\newcommand{\Tr}{\textsf{Treated}}

\newcommand{\Ct}{\textsf{Control}}

\renewcommand{\textrightarrow}{$\rightarrow$}

\colorlet{tableheadcolor}{gray!25} 
\colorlet{tablerowcolor}{gray!5} 

\definecolor{neutralCol}{HTML}{dd1c77}
\definecolor{neutralGreen}{HTML}{31a354}
\definecolor{NewBlue}{HTML}{1879ba}
\definecolor{bleudefrance}{rgb}{0.19, 0.55, 0.91}  
\definecolor{AfTrColor}{HTML}{0868ac}  
\definecolor{BfTrColor}{HTML}{a8ddb5}  

\definecolor{AfCtColor}{HTML}{b10026}  
\definecolor{BfCtColor}{HTML}{fd8d3c}

\graphicspath{ {figures/} }

\newcommand{\itpara}[1]{\vspace{0.3em}\noindent\textit{#1}~}

%% file: 00-body.tex
\input{01-intro}
\input{02-related-work}

\input{02.5-data}
\input{03-methods}
\input{04-results}

\input{05-discussion}
\input{06-limitations}
\input{07-conclusion}

%% file: 01-intro.tex
\section{Introduction}

As the social media ecosystem continues to rapidly expand, platform designers and researchers are experimenting with new models of digital governance~\cite{jhaver2023decentralizing,masnick_2019,2021-modpol}. Recent research has begun extending guiding principles
that could possibly serve such models~\cite{schoenebeck2020drawing,xiao2023rj}. 
This includes rights-based legal approaches, such as international human rights law and American civil rights law~\cite{gorwa2019platform}. 
The HCI community has especially centered around aspirational computer science principles of fairness, accountability, transparency, ethics, and responsibility~\cite{vaccaro2021contestability}. Most famously, a group of human rights organizations, advocates, and academic experts
developed and launched what they termed ``the Santa Clara Principles on Transparency and Accountability in Content Moderation,'' which aim to guide platforms on how to incorporate meaningful transparency and
accountability around moderation of user-generated content~\cite{santaclaraprinciples2020santa}.

Again, empirical research on incorporating meaningful transparency
and how it may benefit users as well as platforms, has begun to
emerge. For example, transparency through removal notification and
providing moderators' reasoning behind content removal has been shown as
one of the key factors in users' perception of the fairness of content
moderation~\cite{jhaver2019survey}. Another study has shown that when offered removal
explanations in any online community, users tend to improve their
posting behavior in that community in the future~\cite{jhaver2019explanations}. Such evidence has been used to motivate platforms, community moderators, and policymakers to continue to push for increased, meaningful transparency in their
moderation practices.

This study seeks to add further empirical evidence to the effects of
offering transparency in content moderation on social media platforms. Specifically, we look at whether such transparency can serve users other than those sanctioned.
Prior research has provided evidence for the educational
benefits of offering removal explanations for users whose content is
removed~\cite{jhaver2019survey,jhaver2019explanations}. However, the effects on \emph{bystanders} who witness the post-removal and the explanation behind it have not been tested. 
Focusing on bystanders allows us to examine the impact of indirect experiences with punishment on users' behavior.
In this research, we ask the question: \emph{Do public removal explanations intended for the sanctioned users influence the posting behavior of bystanders to those explanations?}

We collected a dataset of 85.5M posts from two large Reddit communities, \textit{r/AskReddit} and \textit{r/science}, over the time period Dec 2021--Dec 2022, and developed a computational framework based on causal inference that matched users who witnessed a removal explanation in June 2022 with users who did not witness any explanation. Comparing the post-treatment behavior of these matched groups, we found that exposure to removal explanations significantly boosted the posting activity and interactivity of bystanders as compared to non-bystanders. This shows that the behavioral impacts of moderation transparency on posting volumes are more broadly applicable than previously understood~\cite{jhaver2019explanations,jhaver2019survey}.
Drawing upon this insight, we argue that community managers must invest more time and effort in increasing moderation transparency through explanation messages.
On the other hand, witnessing explanation messages did not significantly enhance the posting quality of bystanders. We speculate on the causes of this empirical insight and offer directions for future research that may help us better understand the role of explanation messages.




%% file: 02-related-work.tex
\section{Background and Related Work}


\subsection{Transparency in Content Moderation}

Moderation systems on social media platforms are designed for governance purposes and often impose measures such as removing content, muting, or banning offenders~\cite{gillespie2018custodians,grimmelmann2015virtues}. These measures are implemented by content moderators, who may either be volunteers among the platform's user base or commercial content moderators hired by the platform~\cite{roberts2019behind,seering2019moderator}. More recently, AI-driven tools have been used to assist in moderation processes~\cite{binns2017like,kiene2020bots,jhaver2023personalizing,horta2023automated}. We focus here on transparency in end-users' experience with moderation processes. \textit{Transparency} implies opening up ``the working procedures not immediately visible to those not directly involved to demonstrate the good working of an institution''~\cite{moser2001open}.

We situate our work within a line of research that examines the impact of content moderation on end-users. Scholars have investigated the impact of both user-level ~\cite{jhaver2021deplatforming,jhaver2019survey,west2017,seering2017} and community-wide sanctions~\cite{chandrasekharan2017you,chandrasekharan2022quarantined}.
This has included studies using a variety of methods, such as interviews~\cite{jhaver2023personalizing}, design workshops~\cite{vaccaro2021contestability}, surveys~\cite{jhaver2019survey,myers2018censored}, and log analyses~\cite{jhaver2021deplatforming,chandrasekharan2022quarantined,chandrasekharan2017you}.
Some studies in this area have also highlighted the benefits of offering moderation explanations to sanctioned users ~\cite{jhaver2019survey,jhaver2019explanations}.
Our focus is also on end-users who witness, although they are not directly affected by, the moderation sanctions.
By doing so, we contribute to building a theory~\cite{kiesler2012} that prescribes to community managers which moderation interventions should be deployed, under what circumstances, and with what expected outcomes.

In examining the complexities of enacting content moderation, researchers have identified several issues regarding transparency in the procedures followed by platforms when applying punitive measures~\cite{renkai2023boiler}. First, the criteria for determining inappropriate content might not be well-established before moderation decisions are made~\cite{seering2020reconsidering}. Legal experts have raised concerns that despite social media platforms publicly sharing their content policies, they often fail to adequately consider the contextual factors surrounding the content, such as its localized meaning and the identities of the speakers and audiences, when evaluating its appropriateness~\cite{wilson2020hate}. 
Second, there are inter-platform differences in how norm violations are conceptualized.
For example, an HCI study comparing the content policies of 15 platforms found a lack of consensus in defining what qualifies as online harassment and how forcefully content deemed as harassment should be moderated~\cite{pater2016}. Consequently, when these vague content policies are implemented for content regulation, it can lead to ambiguity in resolving moderation cases~\cite{wilson2020hate}.
Finally, and most pertinent to our study, communication with end-users regarding moderation decisions is often found to be deficient in details~\cite{myers2018censored,suzor2019we}.

\subsection{Removal Explanations and Bystanders to Norm Violations}

Prior research have emphasized the significance of incorporating moderation notifications and explanations into the design of moderation systems~\cite{jhaver2019explanations,kou2017punishment,renkai2021advertiser,vaccaro2020facebook}.
For example, researchers have shown that when Facebook and Reddit platforms do not inform users about their content removal~\cite{suzor2019we}, users question which platform policy they have violated~\cite{jhaver2019survey,myers2018censored}.
Besides removal notification, users desire a justification for why their posts got removed, deeming it a significant factor in their perception of moderation fairness~\cite{jhaver2019survey}.
Users also express dissatisfaction with the inconsistent punishments meted out to them versus others, leading them to request explanations further~\cite{renkai2022youtube,vaccaro2020facebook}.
Many studies have empirically shown the benefits of offering removal explanations in improving the behavior of moderated users~\cite{jhaver2019survey,jhaver2019explanations,tyler2021social}.
For example, Tyler et al. found that users who were provided education about platform rules in the week following their post removal were less likely to post new violating content~\cite{tyler2021social}.
We extend this research by investigating the utility of explanations in influencing the behavior of bystanders.

Curiously, Reddit moderators offer explanations publicly by commenting on the removed submission. While this is not the sole communication mode---indeed, many moderators privately message users to inform them about moderation---prior research has argued that public explanations serve to enhance broader transparency efforts~\cite{jhaver2019survey,jhaver2019explanations}. On Reddit, users already engaging with a post retain access to it even after it is removed from the main subreddit; in this sense, removed submissions are not really \textit{removed}, just hidden from the public view. By publicly explaining the reason behind post removal, explanation comments serve users who stumble upon it or are already engaged. 

We extend prior inquiries into using Deterrence Theory~\cite{stafford1993deter} to evaluate the impact of punishments on deterring inappropriate behaviors online~\cite{seering2017,gillett2023hack}.
Deterrence Theory makes a distinction between general and specific deterrence --- specific deterrence refers to the effect of punitive measures on individuals subjected to them, while general deterrence pertains to the impact of the potential threat of such measures on uninvolved observers.
By focusing on bystanders, we examine the effects of generalized deterrence in shaping user behavior. 
Seering et al. showed that banning any type of behavior on Twitch significantly reduced the frequency of that behavior in subsequent messages posted by others~\cite{seering2017}. 
Building upon this, we examine whether clarifying which aspects of submissions prompt sanctions via explanation messages influences observers' subsequent actions.

Encouraging voluntary compliance with behavioral norms in a community requires that community members know the norms and be aware of them when being active within the community. Kiesler et al.~\citep{kiesler2012} argue that people learn the community norms in three ways: (1) observing other people's behavior and its consequences, (2) seeing codes of conduct, and (3) behaving and directly receiving feedback. 
Prior research has demonstrated the importance of users seeing codes of conduct~\cite{matias2016posting} and directly receiving feedback in improving their subsequent behavior~\cite{jhaver2019explanations,tyler2021social}. 
We focus here on establishing the utility of bystanders observing other people's norm violations and the resulting consequences.

In terms of reducing the posting of norm-violating content, some research has focused on the roles bystanders can play in the context of online harassment. 
Blackwell et al. found that labeling a variety of technology-enabled abusive experiences as `online harassment' helps bystanders \textit{understand} the breadth and depth of this problem~\cite{blackwell2017classification}. 
Further, designs that motivate bystander intervention discourage harassment through normative enforcement~\cite{blackwell2018online}. 
Taylor et al.~\cite{taylor2019empathy} additionally found that design solutions that encourage empathy and accountability can promote bystander intervention to cyberbullying. 
Extending this line of research to a broader range of norm violations, we analyze how bystanders are affected by their exposure to post-removal explanations. 


%% file: 02.5-data.tex
\section{Data and Methods}
\subsection{Study Design and Rationale}
We conducted an observational study to examine the effects of witnessing post-removal explanations on Reddit.
Prior HCI and CSCW research has recognized that observational analyses of social media data can serve as a valuable tool for understanding society and evaluating changes in users' behavior, especially regarding their use of social network sites~\cite{sleeper2015goals}. 
Regarding our study's context, empirical research on the effects of various content moderation interventions has often deployed observational analyses of social media logs~\cite{seering2017,chandrasekharan2022quarantined,saha2019prevalence,chandrasekharan2017you}.
Similar to our work, such research has primarily examined behavior patterns over more extended timeframes, typically spanning months~\cite{jhaver2019explanations,jhaver2021deplatforming,horta2021platform}.

Examining the impact of an intervention, whether internal or external, is best studied through causal inference approaches, such as  randomized controlled trials (RCTs).
However, these approaches have certain limitations. 
First, experimental studies requiring participant consent can be constrained by concerns about the observer effect~\cite{mccambridge2014systematic}---that individuals might alter their typical behavior when they are aware of being monitored or observed.
Second, conducting experimental research without participants' awareness is considered unethical, especially within the human-centered research paradigm~\cite{jouhki2016facebook,metcalf2016human}. 
Finally, conducting experiments without prior awareness of their potential impact on participants can lead to long-term adverse consequences for both platforms and individuals.

As a result, observational studies can serve as a viable alternative in situations where experimental approaches may not be feasible or ethical. While observational studies may not provide true causality, they are structured to minimize confounds and investigate longitudinal data, offering stronger evidence than basic correlational analyses~\cite{imbens2015causal}. 
Recently, there has been growing interest in these types of studies within the fields of HCI and behavioral science, including those analyzing social media data ~\cite{dechoudhury2016,kiciman2018using,olteanu2017distilling,sadilek2012modeling,saha2019social,zhang2020conv,saha2021advertiming,keith2020text}. 
Significantly, the research conducted by~\citeauthor{saha2018social} prompted us to operationalize metrics for assessing social media behavior, including factors like activity and interactivity~\cite{saha2018social}. 

Given the above considerations, we drew on quasi-experimental approaches to observational data. We adopted a causal-inference approach based on the potential outcomes framework proposed by~\citet{rubin2005causal}.~\autoref{fig:causalSchematic} shows a schematic figure of our approach. This approach simulates an experimental setting by matching individuals (\Tr{} and \Ct{}) on several covariates~\cite{imbens2015causal}. For a given treatment, $T$, two potential outcomes are compared: (1) when a user is exposed to $T$ ($T=1$), and (2) when a user is not exposed to $T$ ($T=0$). Because it is impossible to obtain both kinds of outcomes simultaneously for the same user, this framework estimates the missing counterfactual for a user based on the outcomes of a matched user---another user with similar covariates (attributes and behaviors) but not exposed to $T$. Our work drew motivation from prior works that adopted similar causal-inference approaches on social media data~\cite{saha2019social,chandrasekharan2017you,kiciman2018using}.

\begin{figure*}
\centering
 \begin{subfigure}[b]{0,8\columnwidth}
    \centering
    \includegraphics[width=\columnwidth]{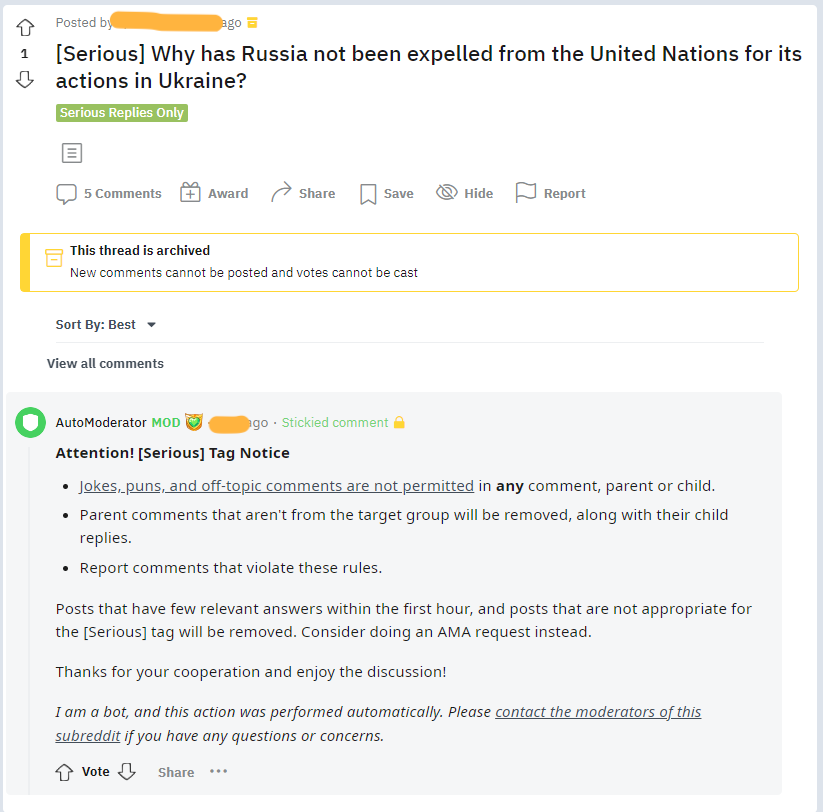}
    \caption{}
    \end{subfigure}\hfill
 \begin{subfigure}[b]{\columnwidth}
    \centering
    \includegraphics[width=\columnwidth]{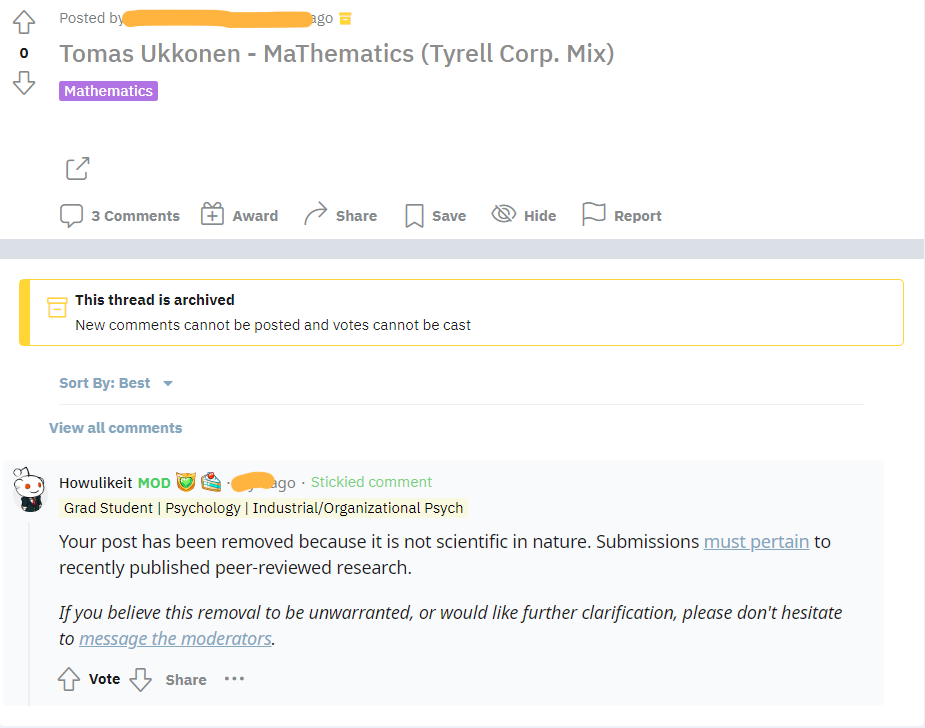}
    \caption{}
    \end{subfigure}\hfill
\caption{Examples of post-removals and explanations by a moderator on (a) \textit{r/AskReddit} (here, the explanation is provided by the Automoderator), and (b) \textit{r/science} (here, the explanation is provided by a human moderator).}
\label{fig:exampleRemovals}
\end{figure*}

\subsection{Choice of Subreddits}
This paper focuses on two major subreddits, \textit{r/AskReddit} (43M members) and \textit{r/science} (31M members).
r/AskReddit is a community focused on asking and answering questions that elicit thought-provoking discussions, offer light entertainment, and help users learn more about their fellow community members.\footnote{\url{https://www.reddit.com/r/AskReddit/wiki/index}}
r/science is a science news and discussion community where users post links to research papers or reputable news items representing recent scientific research, and engage in science communication~\cite{jones2019rscience}.

We analyze these two communities for two main reasons. First, due to their importance --- they are among the largest and most active Reddit communities and have impacted society at large, e.g., through widespread sharing of personal experiences, expert testimony, and science communication on a range of topics~\cite{lanius2019torment,jones2019rscience}.
Second, both communities have a mature moderation approach --- they have been active for more than 15 years, and have a well-described set of posting guidelines and dozens of active moderators. This made it more likely that their approach to offering removal explanations would deliver messages appropriate for our study.

~\autoref{fig:exampleRemovals} shows example post-removals on the subreddits. We downloaded the data from these subreddits over 13 months between 01 December 2021--31 December 2022, using the \textit{pushshift.io} service. 

\begin{table}[t]
  \begin{center}
\sffamily
\footnotesize
\centering
\caption{Summary statistics of the Reddit dataset.}
\begin{tabular}{crr} 
 \textbf{Subreddit} & \textbf{No. Submissions} & \textbf{No. Comments} \\
 \bottomrule
 \textit{r/Askreddit} & 287,954 & 5,358,662 \\ 
 \textit{r/science} & 2,453 & 175,007 \\
 \bottomrule
\end{tabular}
\label{table:datastats}
\end{center}
\end{table}

We iterated through this dataset, decompressing and decoding it in smaller chunks, and simultaneously storing the readable data into SQLite database tables. We queried the database to access the data for the ensuing analyses in the paper.~\autoref{table:datastats} summarizes the data (submissions and comments) collected for our study. Note that we use the term \textit{post} to indicate posting activity in the form of either submissions or comments; therefore, for any given period $\mathtt{T}$, $\mathtt{N_p (T) = N_s (T) + N_c (T)}$ (where $\mathtt{N_p}$, $\mathtt{N_s}$, and $\mathtt{N_c}$ denote the number of posts, submissions, and comments respectively).






\subsection{Defining \Tr{} and \Ct{} Users}
Our study employed a causal-inference framework, drawing on similar approaches in prior research~\cite{chandrasekharan2017you,saha2018social,kiciman2018using}. 
For this purpose, we defined \textit{treatment} as exposure to post-removal explanation(s). 
Within our study period of 13 months, we considered the period between 01-30 June 2022, as our \textit{treatment} period, i.e., we focused on explanations provided in this one month and collected six-month pre-treatment and six-month post-treatment period data for our analyses.
We randomly selected June 2022 as our treatment period, following similar selections in prior moderation research~\cite{jhaver2019explanations}.

Our \Tr{} users comprise the ``bystanders'' or the users of a subreddit who witnessed a removal explanation during the treatment period. 
While this set would ideally consist of users who read and comprehended the explanation comment, we did not have access to users' viewing logs.
Therefore, we constituted \Tr{} users by assuming the commenting activity as a proxy for exposure and including users who commented in the discussion thread containing the removal explanation.
On the other hand, \Ct{} users comprise users of the same subreddit who did not comment in any discussion thread containing a removal explanation but posted elsewhere in the same subreddit during the pre-treatment period.  We filtered out the data of any user exposed to post-removal explanations in the period between December 2021--May 2022 to ensure that we examined \Tr{} users subjected to \textit{treatment} in June 2022.


\subsection{Gathering Post Removal Explanations Data}
We obtained a list of 95 phrases indicating post-removal explanations from prior work by~\citet{jhaver2019explanations}. We used these phrases to query our database to collect 
all the removal explanations in our defined \textit{treatment} period. 
Specifically, we queried the created database to retrieve the stickied\footnote{Removal explanations are usually stickied, i.e., locked to appear as the top comment in the discussion thread.} comments made by moderators within the \textit{treatment} period and containing any of the above phrases. 
We obtained 
257 removal explanations on \textit{r/AskReddit} and 379 such removal explanations on \textit{r/science}.
Focusing on the discussion threads of each of these removal explanations, we next collected the information of the commenters, who are the bystanders or \Tr{} users in our study. 
In some threads, the removed submission's author also posted a comment in the discussion thread; we did not include such submission authors in the \Tr{} users groups because our analysis centers on bystanders, not moderated users.

We obtained the timeline of posts made by the \Tr{} users in the corresponding subreddit during the study period. 
For each subreddit, we also curated a list of  \Ct{} users: this constituted users who were not \Tr{} users, and who were not exposed to any post-removal explanations in the pre-treatment period. The \Tr{} users were assigned with a \textit{treatment date} on their first occurrence of witnessing a post-removal explanation during our \textit{treatment period}. On the other hand, because the \Ct{} users did not have any \textit{treatment date} per se, we simulated a set of \textit{placebo dates} from the set of all possible \text{treatment dates} within the subreddit, such that the distributions of \textit{placebo dates} and \textit{treatment dates} were statistically similar. Then, each \Ct{} user was randomly assigned a \textit{placebo date} from the set of \textit{placebo dates}. For easier readability, any following reference to pre-treatment and post-treatment surrounds \textit{treatment date} for a \Tr{} user, and \textit{placebo date} for a \Ct{} user.

%% file: 03-methods.tex
\subsection{Matching for Causal-Inference}

\begin{figure}
\centering
\includegraphics[width=\columnwidth]{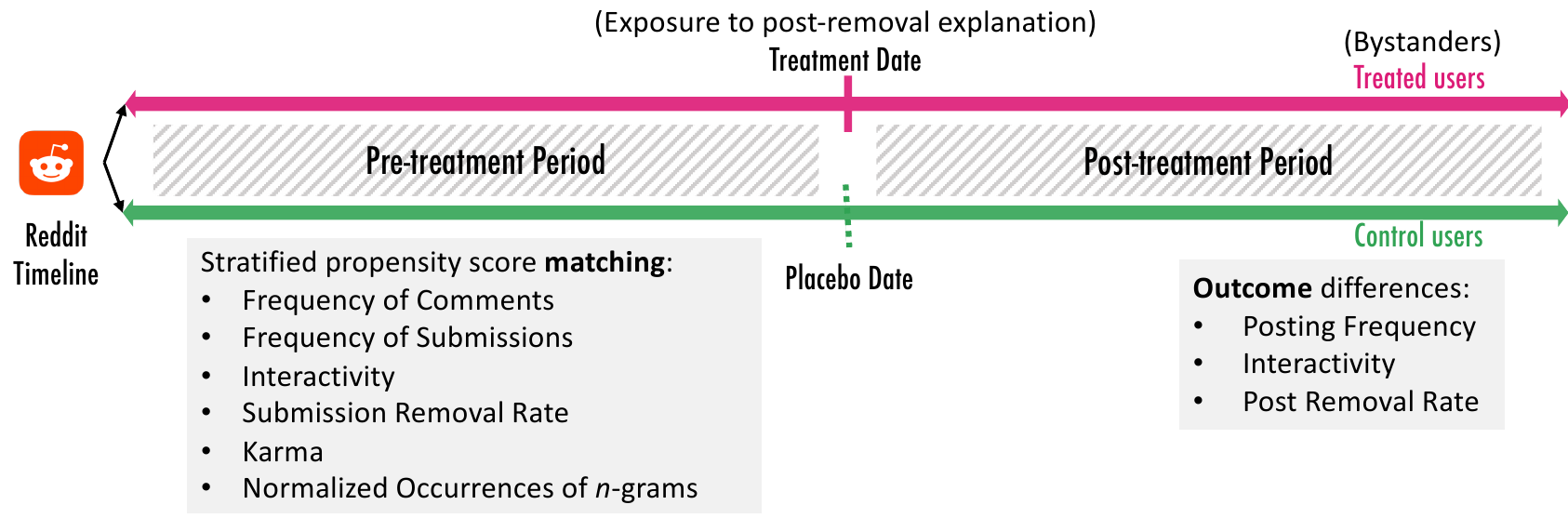}
\vspace{-1em}
\caption{A schematic figure showing our causal-inference approach to analyze users' Reddit timeline.}
\vspace{-1em}
\label{fig:causalSchematic}
\end{figure}

\subsubsection{Covariates for Matching} We operationalized a number of covariates that we would use for matching the \Tr{} and \Ct{} users, motivated from prior work~\cite{kiciman2018using,saha2018social,de2016discovering,jhaver2019explanations,chandrasekharan2017you,saha2020causal}, as listed below. Each covariate was measured using the data in the user's pre-treatment history.

    
    
    \itpara{Frequency of Comments}: The normalized quantity of comments per day, as also used in prior work~\cite{saha2018social,chandrasekharan2017you}. 
    
    \itpara{Frequency of Submissions}: The normalized quantity of submissions per day, as also used in prior work~\cite{saha2018social,chandrasekharan2017you}. 

    \itpara{User Interactivity}: The ratio of number of comments to the total number of posts, as also used in prior work~\cite{saha2018social,saha2020causal}.    
    
    \itpara{Submission Removal Rate}: The ratio of removal submissions to total submissions posted by the user, as also used in prior work~\cite{jhaver2019explanations}.
    
    \itpara{Karma}: Average karma across the comments and submissions made by the user, as also used in prior work~\cite{chandrasekharan2017you,saha2018social}.
    
    \itpara{Normalized $n$-grams}: The normalized occurrences of the top 1000 $n$-grams ($n=1, 2$),
     as also used in prior work~\cite{saha2019social,de2016discovering}.

\subsubsection{Stratified Propensity Score Matching} As mentioned above, we used matching to find pairs (generalizable to groups) of \Tr{} and \Ct{} users with statistically similar covariates. We adopted the propensity score matching approach that matches users based on propensity scores, which is essentially a user's \textit{likelihood} of receiving the treatment. However, exact one-to-one propensity score matching can suffer from biases~\cite{king2019propensity}. Therefore, motivated by prior work~\cite{kiciman2018using,saha2021advertiming,yuan2023mental}, we adopted \textit{stratified propensity score matching} that can balance the bias-variance tradeoff of either too biased (one-to-one match) or too variant (unmatched) data comparisons. In a stratified matching approach, users with similar propensity scores are grouped into strata. Hence, every stratum consists of users with similar covariates~\cite{kiciman2018using}. Through this approach, we isolated and estimated treatment effects within each stratum.

For the above matching, we computed the propensity scores by building a logistic regression model with the covariates as independent variables and a user's binary treatment score (1 for \Tr{} users and 0 for \Ct{} users) as dependent variable. We segregated the distribution of propensity scores into 200 strata of equal width. To ensure that our causal analysis was restricted to a sufficient number of similar users, we discarded strata with less than 10 \Tr{} and 10 \Ct{} users. This led to a final matched dataset of 50 strata (4,842 \Tr{} users and 146,922 \Ct{} users) in \textit{r/AskReddit} and 33 strata (4,890 \Tr{} users and 176,324 \Ct{} users) in \textit{r/science}.

\subsection{Measuring Treatment Effects}
After matching the \Tr{} and \Ct{} users, we measured the differences in the post-treatment behaviors of the users. For this, we operationalized three outcomes---1) \textit{Frequency of posting}, 2) \textit{Interactivity}, and 3) \textit{Submission Removal Rate} for the users in the post-treatment period. 
We draw on the difference in differences approach in causal-inference~\cite{goodman2021difference} and prior work using these approaches on social media~\cite{chandrasekharan2017you,de2016discovering,saha2020causal,saha2019social}, to calculate the average treatment effect (ATE) as the average of the difference of changes in the \Tr{} users and the \Ct{} users per stratum. In addition, we obtained the effect size (Cohen's $d$) and evaluated statistical significance in differences using relative $t$-tests. We conducted Kolmogorov-Smirnov ($KS$) test to evaluate the differences in the distributions of the \Tr{} and \Ct{} groups' outcomes. 






%% file: 04-results.tex
\section{Results}
\autoref{tab:result} summarizes our observations of the differences in the post-treatment outcomes in our study. We describe our findings below:

\begin{table}[t!]
  \begin{center}
\centering
\sffamily
\footnotesize
\caption{Summary of changes in outcomes for the \Tr{} and \Ct{} individuals. We report average treatment effect (ATE), effect size (Cohen's $d$), relative $t$-test, and KS-test statistics (* $p<0.01$, ** $p<0.001$, *** $p<0.0001$.}
\label{tab:result}
\begin{tabular}{lrrr@{}lr@{}l}
\textbf{Outcome} & \textbf{ATE} & \textbf{Cohen's $d$} & \multicolumn{2}{c}{\textbf{$t$-test}} &\multicolumn{2}{c}{\textbf{$KS$-test}} \\ 
\toprule
\rowcollight \multicolumn{7}{c}{\textbf{\textit{r/AskReddit}}}\\
Posting Frequency & 0.453 & 0.807 & 6.589 & *** & 0.640 & ***\\
Interactivity & 0.193 & 2.392 & 12.233 & *** & 0.960 & ***\\
Post Removal Rate & 0.000 & 0.005 & 0.024 & & 0.200 & \\
\rowcollight \multicolumn{7}{c}{\textbf{\textit{r/science}}}\\
Posting Frequency & 0.025 & 1.075 & 8.890 & *** & 0.515 & ***\\
Interactivity & 0.216 & 1.445 & 17.469 & *** & 0.879 & ***\\
Post Removal Rate & 0.001 & 0.007 & 0.177 & & 0.303 & \\
\bottomrule
\end{tabular}
\end{center}
\end{table}

\textit{Posting Frequency.} We find significant differences in the posting frequency of \Tr{} and matched \Ct{} individuals. On \textit{r/AskReddit}, the ATE is 0.453, which can be roughly interpreted as the treatment increases the frequency of posts by 1 for about 45.3\% of the individuals. We see a high effect size (0.807) and significant differences as per $t$-test and $KS$-test ($p<0.0001$). We also see convergent findings in \textit{r/science} with an ATE of 0.025, Cohen's $d$ of 1.075, and significant differences as per $t$-test and $KS$-test ($p<0.0001$). Higher posting frequency indicates that the \Tr{} users (bystanders) became more active in the subreddits after witnessing the post-removal explanations. This measure is an indicator of positive community behavior~\cite{saha2018social}.

\textit{Interactivity.} Similar to the above, we find significant differences in the interactivity of \Tr{} and \Ct{} individuals. On \textit{r/AskReddit} we find an ATE of 0.193 and Cohen's $d$ of 2.392, along with statistical significance in differences as per $t$-test and $KS$-test ($p$<0.0001). Likewise, on \textit{r/science}, ATE on interactivity is 0.216, Cohen's $d$ is 1.445, and $t$-test and $KS$-tests reveal statistical significance ($p$<0.0001). In addition to higher posting frequency, higher interactivity indicates that the \Tr{} users not only created more new submissions but also replied more to others' threads---an important factor for enhancing online community engagement~\cite{saha2018social,saha2020causal}. This suggests that post-removal explanations can potentially enhance community engagement and, subsequently, the sustainability and growth of a community with member activity.

\textit{Post Removal Rate.} Interestingly, we find no significant effects on the post-removal rates. That is, we do not have conclusive evidence if the posting quality improved (or worsened) for the \Tr{} users. 

%% file: 05-discussion.tex
\section{Discussion}

\subsection{Implications}
Online communities rely on content generated by users, but inappropriate posts can detract from the quality of the user experience. Consequently, moderation systems typically aim to boost the overall volume of contributions while reducing the need for post-removals~\cite{grimmelmann2015virtues,kiesler2012}. Our analysis in this paper examined the behavioral impact of offering moderation explanations on bystanders over two dimensions---their future posting activity and the frequency of their future post-removals. 
Our results represent the impact of generalized deterrence --- the indirect experience with punishment.
Consistent with prior applications of Deterrence Theory in online platforms~\cite{seering2017,gillett2023hack}, we show that understanding the reasons for sanctions on another user significantly shapes observers' behaviors.
In this section, we examine the implications of our findings for moderators, site administrators, designers, and future research.

\subsubsection{Removal Explanations Help Boost Posting Frequency.}
We found that on both \textit{r/AskReddit} and \textit{r/science}, users who got exposed to removal explanations directed at moderated others significantly increased their posting activity as compared to users who did not witness any explanations. It could be that seeing explanation messages indicated to bystanders that the community is well-moderated. This, in turn, could have enhanced their inclination to be active within the community.

We note that this result contrasts~\citeauthor{jhaver2019explanations}' findings for moderated users---exposure to removal explanations reduced these users' future posting activity~\cite{jhaver2019explanations}. One reason for this could be that users who suffer moderation may find it more difficult to accept the justification for their post-removals than other bystanders.
Prior work has often grappled with the tradeoffs of moderation actions reducing posting traffic at the cost of improving posting quality~\cite{jhaver2019explanations,jhaver2019survey,jhaver2021deplatforming,horta2021platform}.
However, as this study's framing highlights, for any given removed submission, there is only one moderated user but potentially many more bystanders. Thus, our results suggest that providing explanation messages may boost the overall posting frequency in a community.
This empirical insight offers a powerful incentive to community managers considering the deployment of explanation messages.

\subsubsection{Removal explanations help increase community engagement. }
We found that exposure to others' explanation messages increases the posting interactivity. That is, bystanders' comments constitute a greater proportion of their posting volume after the treatment. Prior research has shown that this metric is an important factor in community engagement~\cite{saha2018social}.
Therefore, this finding suggests that observing the reasoned explanation for post removals can inform bystanders why certain types of posts are unacceptable in the community, help them learn its accepted norms~\cite{chandrasekharan2018norms}, and thereby increase their confidence in instituting a deeper engagement with the community. This further demonstrates the utility of offering post-removal explanations.

Another explanation for this finding is that users perceive moderators attend to and regulate inappropriate submissions more than inappropriate comments. This perception may incline them to engage more in posting comments than submissions in an effort to avoid experiencing post removals themselves.
As prior research shows, users often develop ``folk theories'' of content moderation processes in order to make sense of them~\cite{eslami2015always,jhaver2019survey}. 
Going forward, qualitative studies could inquire whether the posting activity of users is shaped by their folk theories of where the content moderation efforts are focused.

\subsubsection{Removal explanations do not impact post removals.}
Our analysis shows that removal explanations do not significantly impact the future post-removals of bystanders. This contrasts previous results for moderated users: \citeauthor{jhaver2019explanations} showed that offering removal explanations reduced the future post-removals of moderated users~\cite{jhaver2019explanations}.
This suggests that explanation messages boost the posting \textit{quality} of moderated users more than bystanders. Why is this the case? One reason could be that having experienced post removal, moderated users may be likelier to attend to \textit{all} community guidelines before posting their next submissions. On the other hand, witnessing a removal explanation may not be a strong enough incentive for bystanders to ensure compliance with all community guidelines in their next submissions. 

It is possible that witnessing explanation messages educates bystanders about the violated community norm specific to the corresponding removed post and leads them to avoid the same violation in the future, yet they continue violating other community norms.
While beyond the scope of the current paper, a more granular analysis could examine whether norm-specific learning occurs through removal explanations among bystanders.
Besides, prior research has shown that users often respond to moderation by changing their deviant posting activities to circumvent restrictions, especially when the moderation is automated and reliant on detecting specific keywords~\cite{chancellor2016thyghgapp,jhaver2022filterbuddy}. 
Thus, removal explanations may offer users clues on how to avoid moderation, thereby depicting a paradox of enacting algorithmic transparency~\cite{jhaver2018airbnb}.
Therefore, beyond focusing on post removals, it is important to qualitatively evaluate the extent to which removal explanations prompt users to \textit{sincerely} engage in adhering to the community's expectations.

\subsubsection{Design Implications.} This work bears design implications regarding the positive impacts of enacting transparency in online content moderation. 
The empirical evidence presented here informs community managers to put more effort into providing explanations for their sanctions, and more importantly, make these explanations \textit{publicly visible}, so that they can educate bystanders.
While content moderation actions have proliferated to align with the growing scale of online communities, providing explanations is still not as prevalent. For instance, to conduct this study, we originally started with four large subreddits---we had also collected over $\sim$2M posts from \textit{r/politics} (8.4M users) and \textit{r/technology} (15M users). However, despite being large subreddits and having many moderators, neither of these communities provided any post-removal explanations (which also prevented us from including their data in our analyses). Prior work has noted challenges in providing explanations in all instances, such as moderator fatigue and limitations of automated moderation tools
~\cite{jhaver2019automated,jhaver2019explanations}. 
Many platforms may lack resources to provide moderation explanations.
However, with the advent of generative AI and large-language model-based technologies, it would be interesting to explore the design space of curating automated explanation messages through these emerging technologies. 
Given that user attention is a limited resource~\cite{kiesler2012}, platforms must also negotiate the extent to which norm education through removal explanations intended for others be prioritized in the content shown to the users.

Besides, more research is needed to develop best practices for designing removal explanations in response to specific norm violations and other contextual details. 
The computational framework of our study can be extended to delineate the effects of different features of explanation messages, e.g., explanation length, politeness level, clarifying future graduated sanctions, including face-saving mechanisms~\cite{kiesler2012}. The results of such analyses can inform platform owners and community managers about the suitability of different explanation types. 
Given the inherent connection of explanations to community guidelines, these efforts could also inform the latter's design.
On Reddit, explanations are made publicly visible through a stickied comment on the removed post. The visibility of such explanations can be further enhanced by sending notifications about them to users engaged in the sanctioned discussion threads. 

%% file: 06-limitations.tex
\subsection{Limitations and Future Directions}

Our analyses focused on two large Reddit communities. Therefore, our results are most readily applicable to other subreddits of similar size. Future analyses would benefit from investigating the circumstances under which these results replicate (or do not) on other platforms and communities. The computational framework we have presented here should help such inquiries.
Prior similar efforts on developing extendable computational frameworks for evaluating moderation actions have similarly used data from a limited number of samples~\cite{chandrasekharan2017you,chandrasekharan2022quarantined,trujillo2022donald,jhaver2021deplatforming}.


For this project, we had initially planned a comparative analysis of the effects of human v/s bot explanations on bystanders. However, our data review showed that all r/AskReddit explanations were provided by bots and all r/science explanations by human moderators during the treatment period. Therefore, we could not conduct our planned comparative analysis for either community.
Future work should explore how AI-generated explanations compare to human-offered explanations in influencing bystanders' behavior, extending similar inquiries in prior research~\cite{jhaver2019explanations}.
Additionally, it would be fruitful to investigate how explanation messages shape other aspects of bystanders' behavior, e.g., their use of language, and how other community members respond to them.

Our analysis does not consider the in-situ practical concerns and constraints under which content moderators work~\cite{mcgillicuddy2016,matias2016civic}. 
Therefore, studies that understand how moderators draft, choose, and submit explanation messages, and help create tools that can make the workflow easier would empower moderators to send explanation messages at a better frequency. 

Our data collection constitutes \Tr{} users by including everyone who commented in the discussion thread regardless of when they commented vis-a-vis the explanation message timestamp. This choice was inspired by our observation that Reddit users can access their posting history and may track the discussion long after their comment, especially since Reddit sends users notifications about posting activity in the threads where users contribute.
Since explanation comments are highlighted at the top of the thread regardless of their upvotes and posting time, exposure to them is likely for everyone viewing the thread.
Still, it is possible that some users left the discussion before the explanation message was posted and never returned. Further, some users may have stumbled upon the thread and viewed the explanation message but never commented on the thread.
Therefore, our measure of exposure to explanations is limited by our data access and is not precise. 
Future research may measure this exposure more precisely by tracking users' passive consumption of explanation messages.

%% file: 07-conclusion.tex
\section{Conclusion}
Transparency in communications is a key concern for moderated users~\cite{jhaver2018blocklists,jhaver2022filterbuddy,suzor2019we}.
On the other hand, secretiveness about moderation decisions triggers speculation among users who suspect potential biases~\cite{jhaver2019survey,myers2018censored,jhaver2022filterbuddy}.
In this paper, we focus on one important mode of enacting greater transparency in moderation decisions: publicly visible messaging by moderators that reveals the reasons behind submission removals.
Our analysis shows that witnessing such messages significantly boosts the posting and interactivity levels of bystanders. 
This suggests that adopting an educational approach to content moderation, as opposed to a strictly punitive one, can lead to enhanced community outcomes.